\documentclass[twocolumn,aps,prl,amssymb,amstext,showpacs,nobalancelastpage]{revtex4}
\usepackage{graphicx,float,amssymb,amstext,amsmath}% Include figure files
\usepackage{dcolumn}% Align table columns on decimal point
\usepackage{bm}% bold math

\newcommand{\rem}[1]{}

\newcommand{\R}{{\mathbb R}}

\newcommand{\ui}{\mathrm{i}}

\newcommand{\capsty}{\footnotesize}
\rem{
\newcommand{\Capts}[1]{#1}
\newcommand{\FIGo}[3]{%
\marginpar{ \begin{platz} \label{#1} ~ \end{platz} \vspace*{1.5ex} }}

} % end rem

\newcommand{\Capts}[1]{}
\newcommand{\FIGo}[3]{\begin{figure}%
#3% % comment this line out to remove all figures
\caption[]{\capsty #2}%
\label{#1}%
\end{figure}}
\begin{document}

\title{Reaction dynamics through kinetic transition states}

%\author{\"{U}nver \c{C}ift\c{c}i} %\email{h.waalkens@rug.nl}
%\affiliation{Department of Mathematics\\ Nam\i k Kemal University\\ 59030 Tekirda\u{g}, Turkey}

\author{\"{U}nver \c{C}ift\c{c}i$^{1,2}$ and Holger Waalkens$^2$} %\email{hdullin@usyd.edu.au}
\affiliation{
$^1$Department of Mathematics, Nam\i k Kemal University, 59030 Tekirda\u{g}, Turkey\\
$^2$Johann Bernoulli Institute for Mathematics and Computer Science, University of Groningen,
PO Box 407, 9700 AK Groningen, The Netherlands 
}

\date{\today}

\begin{abstract}

The transformation of a system from one state to another  is often mediated by a  bottleneck in the system's phase space. In chemistry these bottlenecks are known as \emph{transition states} through which the system has to pass in order to evolve from reactants to products. 
The chemical reactions are usually associated with configurational changes where the reactants and products states correspond, e.g., to two different isomers or the undissociated and  dissociated state of a molecule or cluster.
In this letter we report on a new type of bottleneck  which mediates \emph{kinetic} rather than configurational changes.
The phase space structures associated with such \emph{kinetic transition states} and their dynamical implications are discussed for the rotational vibrational motion of a triatomic molecule. An outline of more general related phase space structures with important dynamical implications is given. 

\end{abstract}


\maketitle

%\section{Introduction}

%%%%%%%%%%%%%%%%%%%%%%%%%%%%%%%%%%%%%%%%%%%%%%%%%%%%%%%%%%%%%%%%%%%%%%%%%%%%%%%%%%%%%%%%%%%%%%%%%%%%%%%%%%%%%%
%%%%%%%%%%%%%%%%%%%%%%
\emph{Introduction.--}
Reaction type dynamics is characterized by the property that a system spends long times in one phase space region (the region of the `reactants') and occasionally finds its way  through a bottleneck to another phase space region (the region of the `products').
The main examples are chemical reactions where the bottlenecks are induced by saddle points of the potential energy surface which arise from a Born-Oppenheimer approximation and  determine the interactions between the constituent atoms or molecules involved in the reaction. 
The reactions are then characterized by configurational changes like, e.g.,  in an isomerization or disscociation reaction.
In this case the bottleneck is referred to a  \emph{transition state}. The main idea of \emph{transition state theory} which is the most frequently used approach to compute  reaction rates 
is then to define a surface in the transition state region and compute the rate from the flux through this so called \emph{dividing surface}. For getting the exact rate this way, it is crucial to define the dividing surface in such a way that it is crossed exactly once by reactive trajectories (trajectories that evolve from reactants to products) and not crossed at all by nonreactive trajectories (i.e. trajectories which stay in the reactants or products region). In the 1970's it has been shown by Pechukas, Pollak and others that for systems with two degrees of freedom, such a dividing surface can be constructed from an unstable periodic orbit which gives rise to the so called \emph{periodic orbit dividing surface} (PODS) \cite{PechukasMcLafferty73,PechukasPollak78}. It took several decades to understand how this idea can be generalized to systems with three or more degrees of freedom \cite{Wigginsetal01}. The object which replaces the periodic orbit is a so called \emph{normally hyperbolic invariant manifold} (NHIM) \cite{Wiggins94}. The NHIM does not only allow for the construction of a dividing surface with the desired crossing properties but it also has (like the unstable periodic orbit in the two-degree-of-freedom case) stable and unstable manifolds which have sufficient dimensionality to form separatrices that channel the reactive trajectories from reactants to products  and separate them from the nonreactive ones \cite{Uzeretal02}.
In this paper we show that for systems which are not of type `kinetic-plus-potential', there are saddle type equilibrium points which either instead of the bottlenecks associated with configuration changes induce bottlenecks of a kinetic nature or, more generally, do not even induce bottlenecks but still give rise to NHIMs whose stable and unstable manifolds govern the dynamics near such equilibrium points.

%%%%%%%%%%%%%%%%%%%%%%

\emph{The standard case.--}
For a system of type `kinetic-plus-potential' (so called \emph{natural mechanical systems} \cite{Arnold78}),   saddle points of the potential lead to equilibrium points of Hamilton's equations near which the Hamiltonian can be brought through a suitable choice of canonical coordinates (so called \emph{normal form coordinates} \cite{Waalkensetal08,Uzeretal02}) into the form 
\begin{equation}
H = E_0 + \frac{\lambda}{2}(p_0^2 -q_0^2) + \sum_{k=1}^n \frac{\omega_k}{2}(p_k^2+q_k^2)+\text{h.o.t.}\,, \label{eq:Hamiltonian_standard_case}
\end{equation}
where $f=n+1$ is the number of degrees of freedom and $E_0$ is the energy of the saddle of the potential.
The quadratic part of the Hamiltonian \eqref{eq:Hamiltonian_standard_case}  consists of a parabolic barrier in the first degree of freedom whose steepness is characterized by  $\lambda>0$ (the Lyapunov exponent) and $n$ harmonic oscillators with frequencies $\omega_k>0$, $k=1,\dots,n$.

Let us ignore the higher order terms for a moment and rewrite the energy equation $H=E$ in the form
\begin{equation}\label{eq:energy_eq_standard}
 \frac{\lambda}{2}p_0^2 + \sum_{k=1}^n \frac{\omega_k}{2}(p_k^2+q_k^2)=  E-E_0 + \frac{\lambda}{2}q_0^2\,.
\end{equation}
Then one sees that for $E>E_0$ (i.e. for energies above the energy of the saddle), each fixed value of the \emph{reaction coordinate} $q_0$ defines a $(2f-2)$-dimensional sphere $S^{2f-2}$. The energy surface  thus has the topology of a `spherical cylinder',   $S^{2f-2}\times \R$. 
This cylinder has a wide-narrow-wide geometry where the spheres $S^{2f-2}$ are `smallest' when $q_0=0$
 (see \cite{WaalkensWiggins04} for a more precise statement). In fact the $(2f-2)$-dimensional sphere given by setting $q_0=0$ on the energy surface $H=E>E_0$ defines a dividing surface which separates the energy surface into a reactants region  $q_0<0$ and a products region $q_0>0$. All forward reactive trajectories cross it with $p_0>0$.  All backward reactive trajectories cross it with $p_0<0$.   The condition $p_0=0$ defines a $(2f-3)$-dimensional sphere which  divides the dividing surface into the two hemispheres which have $p_0>0$ and $p_0<0$ and hence can be viewed as the `equator' of the diving surface.  In fact the equator is a \emph{normally hyperbolic invariant manifold} (NHIM) \cite{Wiggins94}. It can be identified with the transition state: a kind of unstable `super molecule' sitting between reactants and products \cite{Pechukas76,Waalkensetal08}. The NHIM has stable and unstable manifolds  $S^{2f-2}\times \R$. They thus have one dimension less than the energy surface $H=E$. They form the phase space conduits for reaction \cite{WaalkensBurbanksWigginsb04}.
The NHIM is of central importance because it dominates the dynamics in the neighborhood of the saddle.
An important aspect of the theory of NHIMs is that they persist if the 
higher order terms in \eqref{eq:Hamiltonian_standard_case}  are taken into account and the energy is not too far above $E_0$ \cite{Wiggins94}. The phase space structures 
above persist accordingly.

%%%%%%%%%%%%%%%%%%%%%%%%%%%%%%%%%%%%%%%%%
%%%%%%%%%%%%%%%%%%%%%%%%%%%%%%%%%%%%%%%%%
\emph{Kinetic transition states.--}
Let us now more generally consider the case of an equilibrium point of  a Hamiltonian system  which  has one pair of real eigenvalues $\pm \lambda$ and $n$ complex conjugate pairs of imaginary eigenvalues $\pm \ui \omega_k$, $k=1,\ldots,n$.
Choosing normal form coordinates near such a so called
 saddle-center-$\cdots$-center equilibrium the Hamiltonian assumes the form \eqref{eq:Hamiltonian_standard_case}. In fact this general case already formed the starting point in  \cite{Wigginsetal01,Uzeretal02} and later works. However all these studies were restricted to the case where the frequencies $\omega_k$ are positive. In this paper we study what happens if we give up this assumption. 
To this end let us  for simplicity consider a system with $f=2$ degrees of freedom with Hamiltonian function 
\begin{equation}
H = E_0 + \frac{\lambda}{2}(p_0^2-q_0^2) + \frac{\omega}{2}(p_1^2+q_1^2)
\end{equation}
with $\lambda>0$ and $\omega<0$.
In his case let us rewrite the energy equation $H=E$  in the form
 \begin{equation}
 \frac{\lambda}{2}q_0^2-\frac{\omega}{2}(p_1^2+q_1^2)=\frac{\lambda}{2}p_0^2+E_0-E\,.
 \end{equation}
 This is the same as \eqref{eq:energy_eq_standard} however with the role of $q_0$ and $p_0$  exchanged and $E-E_0$ replaced by $E_0-E$. Accordingly for $E<E_0$, 
 the right hand side is positive for any $p_0$ and the energy surface has again the structure of a spherical cylinder with a wide-narrow-wide geometry. The crucial difference to the standard case of positive frequencies in Eq.~\eqref{eq:energy_eq_standard} is that the reaction coordinate is  now $p_0$ instead of $q_0$. The bottleneck is thus associated with kinetic rather than configurational changes. As the following example shows such kinetic bottlenecks do indeed exist in many important applications.

%%%%%%%%%%%%%%%%%%%%%%%%%%%%%%%%%%%%%%%%%
%%%%%%%%%%%%%%%%%%%%%%%%%%%%%%%%%%%%%%%%%

%%
%%
\def\figjacobicoord{%
Definition of Jacobi coordinates for a triatomic molecule. 
}
\def\FIGjacobicoord{
\begin{center} 
\includegraphics[width=5cm]{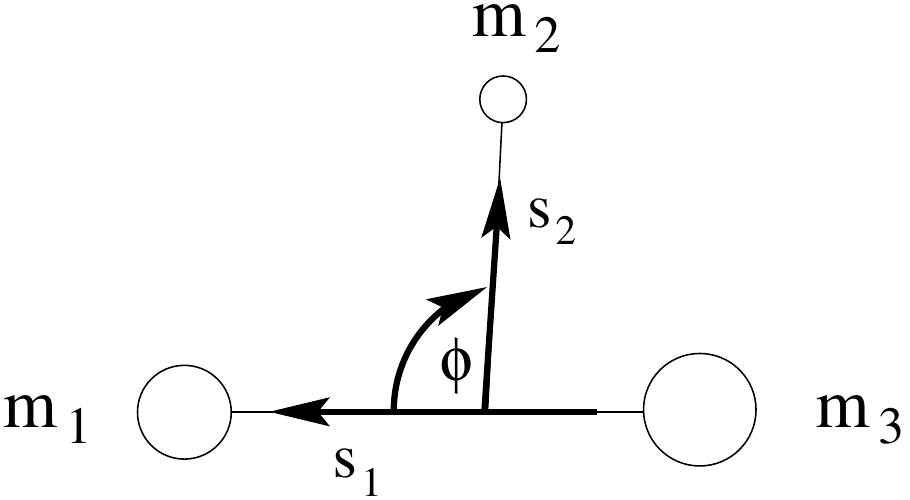}
\end{center}
}
\FIGo{fig:jacobicoord}{\figjacobicoord}{\FIGjacobicoord}

\emph{Example: rotational vibrational motion of triatomic molecules.--}
The Hamiltonian of a triatomic molecule is given by \cite{LittlejohnReinsch97,CiftciWaalkens2012}
\begin{widetext}
\begin{equation}  \label{eq:Hamiltonian_Jacobi}   % \begin{split}
 H=\frac{1}{2} \Big[ \frac{\rho_{1}^{2}+\rho_{2}^{2} \cos^2  \phi}{\rho_{1}^{2}\rho_{2}^{2} \sin^2  \phi}J_{1}^{2}+
\frac{2 \cos \phi}{\rho_{1}^{2} \sin \phi}J_{1}J_{2}+\frac{1}{\rho_{1}^{2}}J_{2}^{2}+
\frac{1}{\rho_{1}^{2}+\rho_{2}^{2}}J_{3}^{2}+p_{1}^{2} +p_{2}^{2}+\frac{\rho_1^2+\rho_2^2}{\rho_1^2 \rho_2^2}(p_{\phi}-\frac{\rho_{2}^{2}}{\rho_{1}^{2}+\rho_{2}^{2}}J_{3})^{2} \Big]+V(\rho_1,\rho_2,\phi)\,. %\end{split}
\end{equation}
\end{widetext}
Here $(\rho_1,\rho_2,\phi)$ are Jacobi coordinates defined as
\begin{equation*}      \label{eq:def_rhophi}        
\rho_1=\|\mathbf{s}_{1}\|\,,\quad \rho_2=\|\mathbf{s}_{2}\|\,, \quad \mathbf{s}_{1}\cdot\mathbf{s}_{2}=\rho_1\rho_2\cos\phi\,,
\end{equation*}
where  $\mathbf{s}_{1}$ and $\mathbf{s}_{2}$ are  the mass-weighted Jacobi vectors (see Fig.~\ref{fig:jacobicoord})
\begin{equation} {\label{eq:def_s1s2}}
\mathbf{s}_{1}=\sqrt{\mu_1}(\mathbf{x}_{1}-\mathbf{x}_{3})\,,\quad
\mathbf{s}_{2}=\sqrt{\mu_2}(\mathbf{x}_{2}-\frac{m_1\mathbf{x}_{1}+m_3\mathbf{x}_{3}}{m_1+m_3}),
\end{equation}
computed from the position vectors  $\mathbf{x}_k$ of the atoms and the reduced masses
\begin{equation}\label{eq:def_reduced_masses}
\mu_1=\frac{m_1m_3}{m_1+m_3} \text{ and } \mu_2=\frac{m_2(m_1+m_3)}{m_1+m_2+m_3}\,.
\end{equation}
The momenta $p_1$, $p_2$ and $p_\phi$ in \eqref{eq:Hamiltonian_Jacobi} are conjugate to $\rho_1$, $\rho_2$ and $\phi$, respectively, and  $\mathbf{J}=(J_1,J_2,J_3)$ is the body angular momentum.
The magnitude $J$ of $\mathbf{J}$ is conserved under the dynamics generated by the Hamiltonian \eqref{eq:Hamiltonian_Jacobi}.

%%%%%
%%% Rigid body case
%%%%%
Let us at first consider a rigid molecule (i.e. the values of $\rho_1$, $\rho_2$ and $\phi$ are fixed). The body fixed frame can then be chosen such that the moment of inertial tensor becomes diagonal with 
 the principal moments inertia $M_1<M_2<M_3$ ordered by magnitude on the diagonal. The Hamiltonian \eqref{eq:Hamiltonian_Jacobi} then reduces to
\begin{equation}  \label{eq:Hamiltonian_rigid}
H = \frac{1}{2} \big( \frac{J^2_1}{M_1} +  \frac{J^2_2}{M_2} + \frac{J^2_3}{M_3}   \big) \,. 
\end{equation}%
As the magnitude $J$ of $\mathbf{J}$ is conserved
the angular momentum sphere $J_1^2+J_2^2+J_3^2=J^2$ can be viewed as the phase space of the rigid molecule \cite{CiftciWaalkens2012}. 
The solution curves of this one-degree-of-freedom system are obtained from the level sets of the Hamiltonian 
$H$ on the angular momentum sphere   (see Fig.~\ref{fig:euler}). 
The Hamiltonian $H$ has local minima at  $(J_1,J_2,J_3)=(0,0,\pm J)$ of energy $J^2/(2M_3)$, local maxima at $(J_1,J_2,J_3)=(\pm J,0,0)$ of energy $J^2/(2M_1)$ and saddles at $(J_1,J_2,J_3)=(0,\pm J,0)$ of energy  $J^2/(2M_2)$. 
These correspond to the center equilibria of stable rotations about the first and third principal axes and the saddle equilibria of unstable rotations about the second principal axis, respectively. 
A possible choice of canonical coordinates $(q,p)$ on the angular momentum sphere is \cite{CiftciWaalkens2012}
\begin{equation} \label{eq:canoncial_coordinates_momentum_sphere}
J_{1} =\sqrt{J^{2}-p^{2}}\cos q,\,
J_{2} =\sqrt{J^{2}-p^{2}}\sin q,\, J_{3} =p \,.
\end{equation}
Since $(J_1,J_2,J_3)=(J,0,0)$ resp. $(q,p)=(0,0)$ is a maximum of the Hamiltonian the normal form of the Hamiltonian at this equilibrium is 
$H=J^2/(2M_1) + \omega (p_0^2+q_0^2)/2 +$h.o.t. with the \emph{negative} frequency $\omega=-\sqrt{M_2M_3(M_2-M_1)(M_3-M_1)}/M_1M_2M_3$.

\def\figeuler{%
Angular momentum sphere with contours of the Hamiltonian in \eqref{eq:Hamiltonian_rigid}. 
}
\def\FIGeuler{
\begin{center} % /home/mazhw/Cpp/ScatteringMonodromy/phase_portrait
\includegraphics[width=6.5cm]{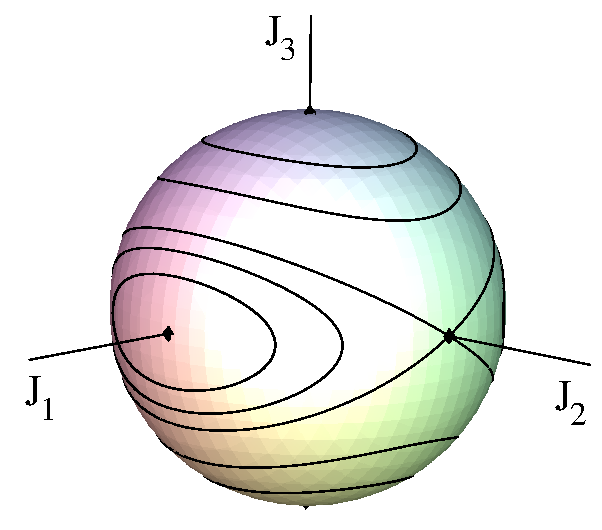} % change to EPS in the end
\end{center}
}
\FIGo{fig:euler}{\figeuler}{\FIGeuler}

Let us now consider a (flexible) triatomic molecule. 
For simplicity, we  freeze $\rho_1$ and $\rho_2$ and consider the two-degree-of-freedom system consisting of pure bending coupled with overall rotations. 
The potential $V$ is then a function of  $\phi$ only and we choose it to be of the form shown in  Fig.~\ref{fig:phipotential}. This potential has two minima at $\phi=0$ and $\phi=\pi$ which correspond
to two different linear isomers which are separated by a barrier at $\phi=\pi/2$. This type  of potential occurs, e.g., in the HCN/CNH isomerization problem \cite{WaalkensBurbanksWigginsb04}.
We consider the equilibrium which for a given magnitude $J$ of the angular momentum arises from the barrier at $\phi=\pi/2$ coupled with rotations about the first principal axis. In the absence of coupling between the bending and rotational degrees of freedom  
we would expect from the discussion of the rigid molecule above that
this equilibrium is  a saddle-center with a negative frequency $\omega<0$. In fact also in the presence of  coupling  between the bending motion and the rotation this remains to be the case (at least for a moderate coupling strength). 
To illustrate the dynamical implication of this equilibrium 
we  use the canonical coordinates $(q,p)$ defined in \eqref{eq:canoncial_coordinates_momentum_sphere} on the angular momentum sphere $J_1^2+J_2^2+J_3^2=J$ and
construct a Poincar{\'e} surface of section  with section condition $q=0\text{ mod }2\pi $, $\dot{q}>0$. Using the canonical pair $(\phi,p_\phi)$ as coordinates on the surface of section we obtain  Fig.~\ref{fig:SOS}.

\def\figphipotential{%
Potential $V$ as a function of the Jacobi coordinate $\phi$.
}
\def\FIGphipotential{
\begin{center} 
\includegraphics[width=6.5cm]{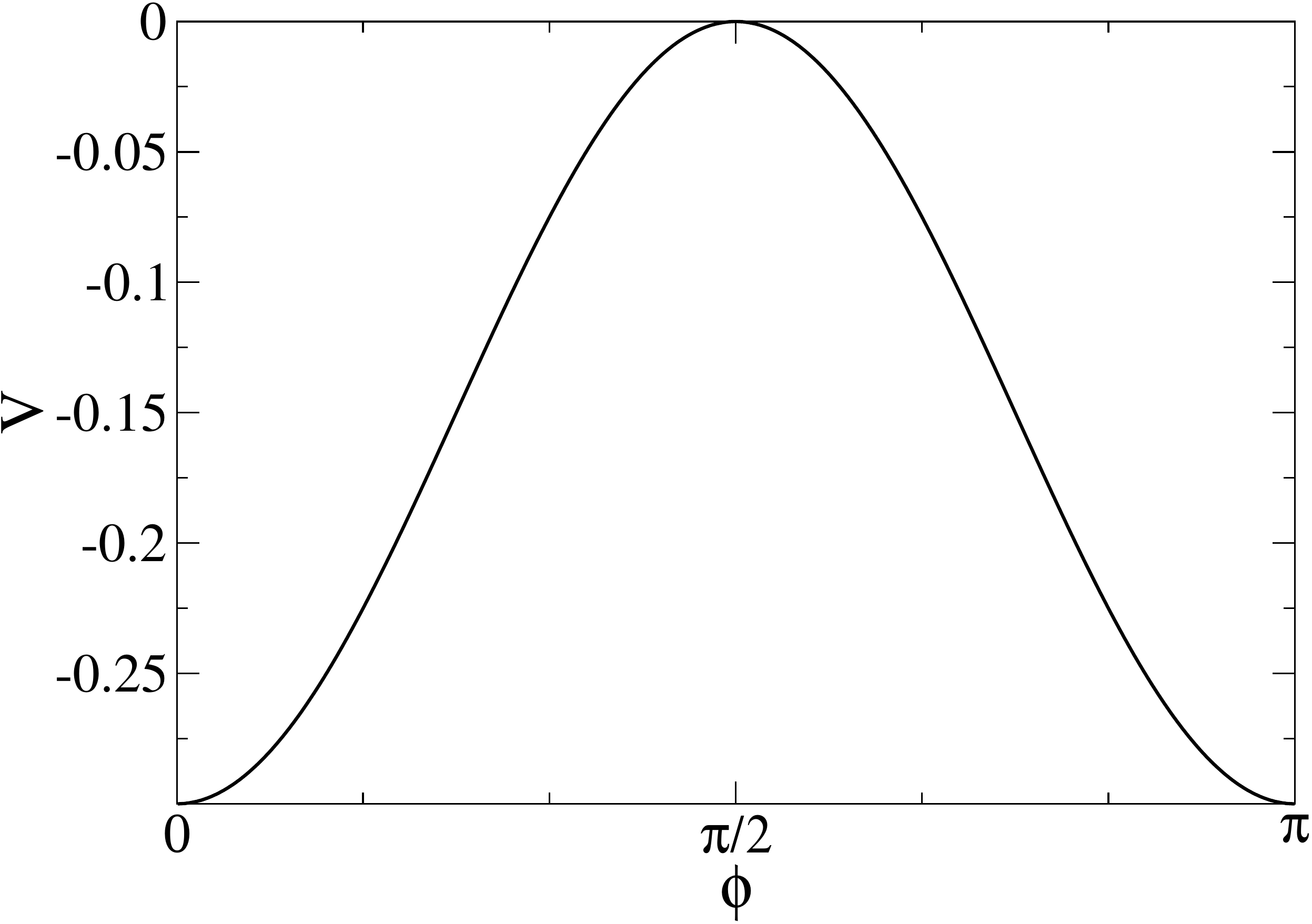}
\end{center}
}
\FIGo{fig:phipotential}{\figphipotential}{\FIGphipotential}

We see that, as expected, there appears to be a barrier associated with the momentum reaction coordinate $p_\phi$. Near $\phi=\pi/2$ no transitions are possible from the  'reactants region' $p_\phi>0$ to the 'products region' $p_\phi<0$ for energies \emph{above} the energy of the saddle whereas for energies \emph{below} the energy of the saddle, transitions are possible.
%Channels between reactants and products: The sign of $p_\phi$ is reversed at $\phi$ near $\phi=0$ or $\phi=\pi$. 
The rotational/vibrational motion of the triatomic molecule with the transition  channel near $\phi=\pi/2$ being closed consists of unhindered bending motion between the two isomers associated with $\phi$ near $0$ and $\pi$, respectively, coupled with rotations.  
For motions with the channel near $\phi=\pi/2$ being  open, $p_\phi$ can switch sign near $\phi=\pi/2$ which corresponds to a trajectory bouncing back   to the isomer it came from rather than switching to the other isomer.  
It is important to note that the saddle-center equilibrium studied here induces a \emph{local} bottleneck for transitions between $p_\phi>0$ and $p_\phi<0$. \emph{Globally} such transitions always occur in the present example when a trajectory passes close to the collinear configuration where $\phi$ is close to $0$ or $\pi$. 
This explains how a single trajectory can contribute points to the lower and the upper half in  Fig.~\ref{fig:SOS}(a) even though the local channel near $\phi=\pi/2$ is closed.

\def\figSOS{%
Poincar{\'e} surface of section for the Hamiltonian \eqref{eq:Hamiltonian_Jacobi} with parameters 
$J= 0.2$, $\rho_1 = 1$ and $\rho_2 = 2$ (see the text). Each picture is generated from a single trajectory of energy
 0.0205625\, in (a) and 0.0197526\, in (b), respectively.
}
\def\FIGSOS{
\begin{center} 
\raisebox{4.5cm}{a)}\includegraphics[width=6.5cm]{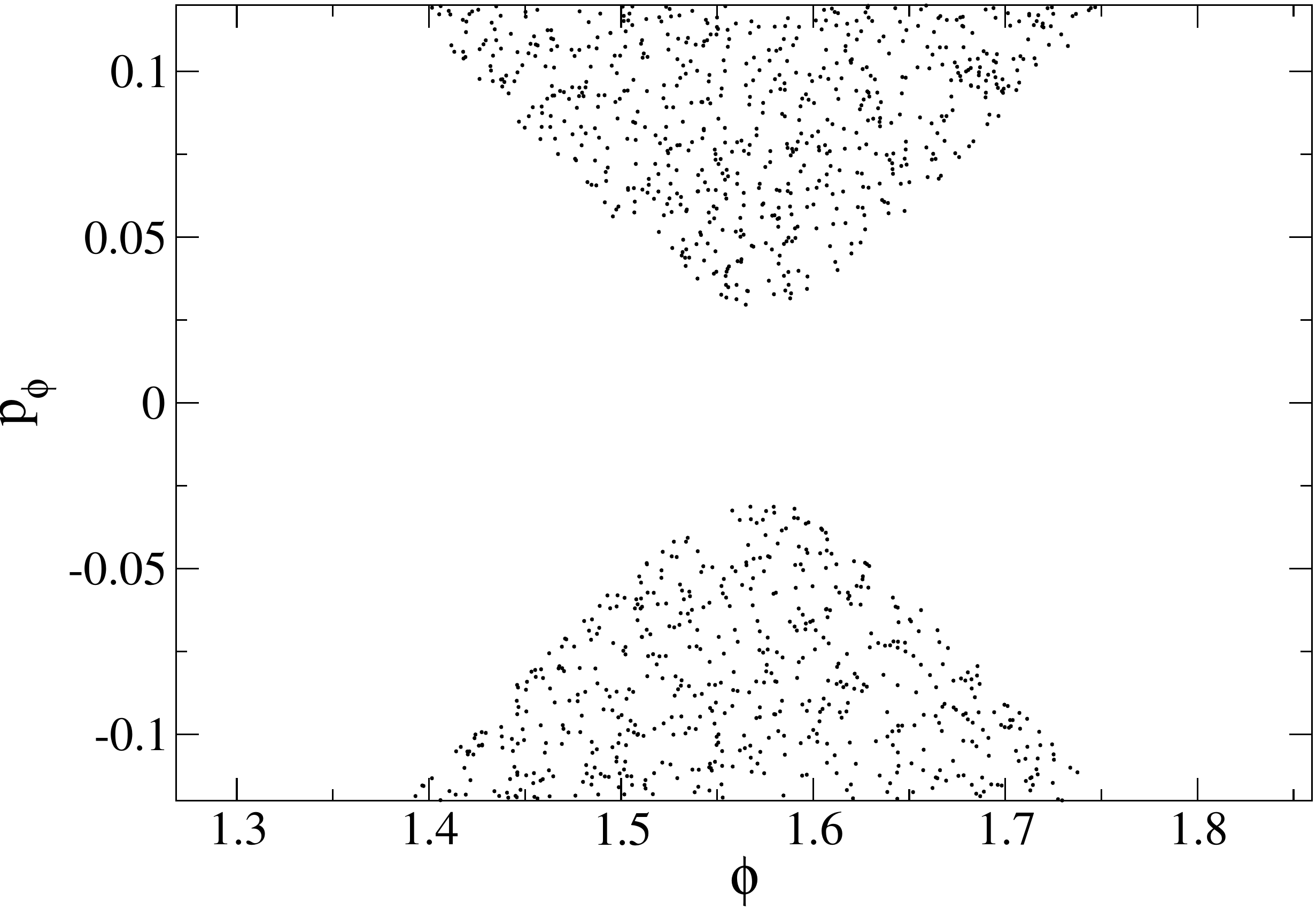}
\raisebox{4.5cm}{b)}\includegraphics[width=6.5cm]{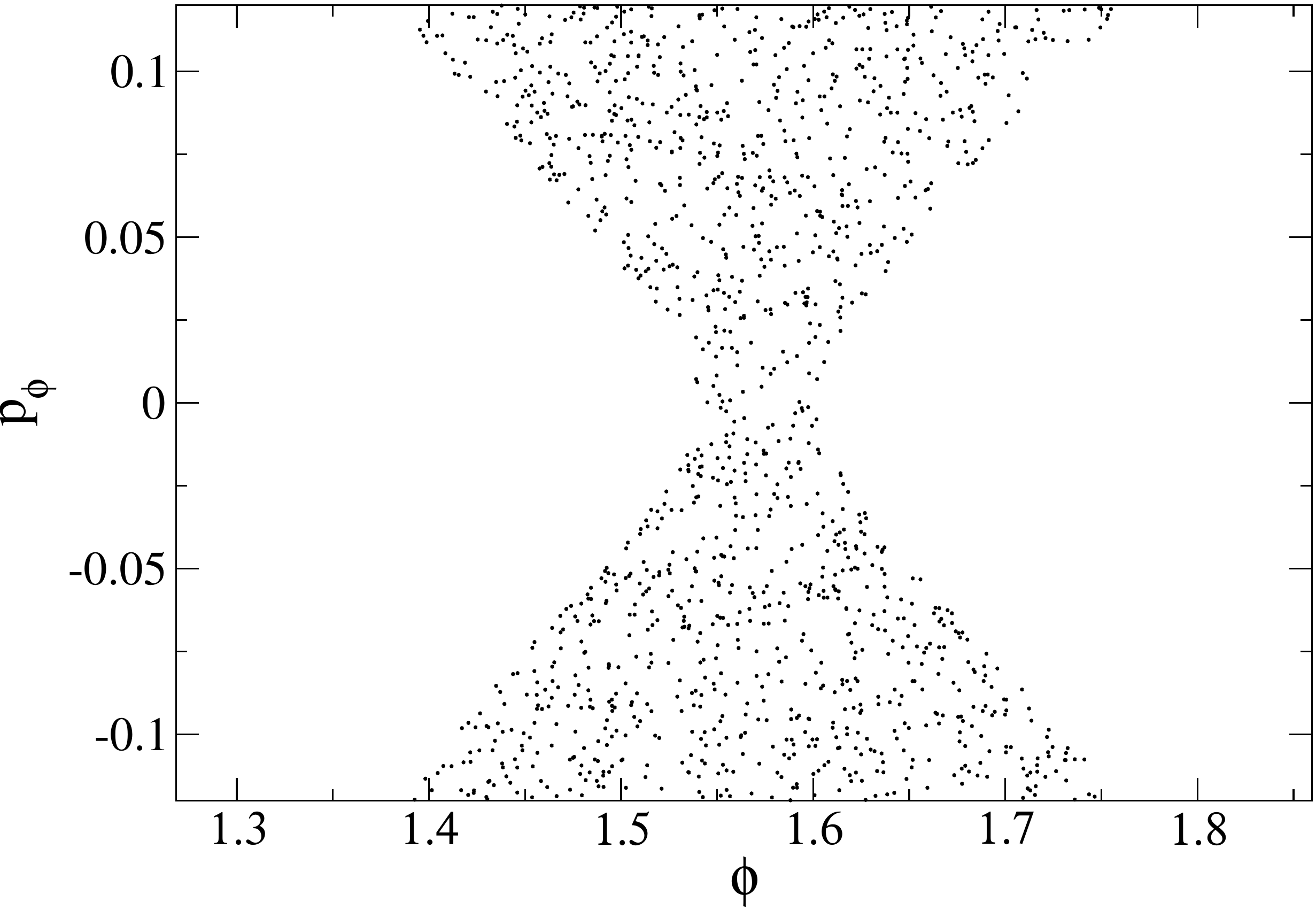}
\end{center}
}
\FIGo{fig:SOS}{\figSOS}{\FIGSOS}

\emph{More generally:  mixed positive and negative frequencies.--}
Near a saddle-center-$\cdots$-center equilibrium 
the Hamiltonian can always be brought into the form \eqref{eq:Hamiltonian_standard_case}. 
The NHIM at energy $E$ is then obtained from the intersection of the center manifold of the equilibrium given by $q_0=p_0=0$ and the energy surface $H=E$, i.e. 
\begin{equation}
 \sum_{k=1}^n \frac{\omega_k}{2}(p_k^2+q_k^2)+\text{h.o.t.} = E - E_0
\end{equation} 
All studies on the geometric theory of reactions  studied so far \cite{TKKBR2005} concern the case of positive frequencies $\omega_k$. 
This corresponds to the Hamiltonian restricted to the center manifold having a minimum. 
 If all frequencies are negative then the  Hamiltonian restricted to the center manifold has a maximum. 
 In this case  the NHIM is again a $(2f-3)$-dimensional sphere. However as opposed to the case of positive frequencies, $p_0$ rather than $q_0$ is the reaction coordinate as discussed in the present example for $f=2$. 
Also the case of mixed signs of the $\omega_k$ occurs in many applications. It, e.g., also occurs
in our example of the rotational-vibrational motion of triatomic molecule if we take the other vibrational degrees of freedom into account which we for simplicity considered to be frozen.
In the case of mixed signs of the $\omega_k$ the NHIM is not a sphere but a non-compact manifold. 
Although no bottleneck in the energy surface is induced in these cases the NHIM has important dynamical implications as it has stable and unstable manifolds which are of one dimension less than the dimension of the energy surface and hence form impenetrable barriers. 
This bears some similarities to the case of 
non-compact NHIM's that have recently  been considered for rank 2 and higher rank saddles \cite{EzraWiggins09,Halleretal11}. The case of mixed positive and negative frequencies 
will form an interesting direction for future research.

%%%%%%%%%%%%%%%%%%%%%%
% \section*{Acknowledgments}

%\bibliographystyle{unsrt}
%\bibliography{chembib}

\begin{thebibliography}{10}

\bibitem{PechukasMcLafferty73}
P.~Pechukas and F.~J. McLafferty.
\newblock On transition-state theory and the classical mechanics of collinear
  collisions.
\newblock {\em J. Chem. Phys.}, 58:1622--1625, 1973.

\bibitem{PechukasPollak78}
P.~Pechukas and E.~Pollak.
\newblock Transition states, trapped trajectories, and classical bound states
  embedded in the continuum.
\newblock {\em J. Chem. Phys.}, 69:1218--1226, 1978.

\bibitem{Wigginsetal01}
S.~Wiggins, L.~Wiesenfeld, C.~Jaff{\'e}, and T.~Uzer.
\newblock Impenetrable {B}arriers in {P}hase-{S}pace.
\newblock {\em Phys. Rev. Lett.}, 86:5478--5481, 2001.

\bibitem{Wiggins94}
S.~Wiggins.
\newblock {\em Normally {H}yperbolic {I}nvariant {M}anifolds in {D}ynamical
  {S}ystems}.
\newblock Springer, Berlin, 1994.

\bibitem{Uzeretal02}
T.~Uzer, C.~Jaff{\'e}, J.~Palaci{\'a}n, P.~Yanguas, and S.~Wiggins.
\newblock The geometry of reaction dynamics.
\newblock {\em Nonlinearity}, 15:957--992, 2002.

\bibitem{Arnold78}
V.~I. Arnold.
\newblock {\em Mathematical Methods of Classical Mechanics}, volume~60 of {\em
  Graduate Texts in Mathematics}.
\newblock Springer, Berlin, 1978.

\bibitem{Waalkensetal08}
H.~Waalkens, R.~Schubert, and S.~Wiggins.
\newblock Wigner's dynamical transition state theory in phase space: classical
  and quantum.
\newblock {\em Nonlinearity}, 21(1):R1--R118, 2008.

\bibitem{WaalkensWiggins04}
H.~Waalkens and S.~Wiggins.
\newblock Direct construction of a dividing surface of minimal flux for
  multi-degree-of-freedom systems that cannot be recrossed.
\newblock {\em J. Phys. A}, 37:L435--L445, 2004.

\bibitem{Pechukas76}
P.~Pechukas.
\newblock In W.~H. Miller, editor, {\em Dynamics of {M}olecular {C}ollisions}.
  Plenum Press, New York, 1976.

\bibitem{WaalkensBurbanksWigginsb04}
H.~Waalkens, A.~Burbanks, and S.~Wiggins.
\newblock Phase space conduits for reaction in multidimensional systems: {HCN}
  isomerization in three dimensions.
\newblock {\em J. Chem. Phys.}, 121(13):6207--6225, 2004.

\bibitem{LittlejohnReinsch97}
R.~G. Littlejohn and M.~Reinsch.
\newblock Gauge fields in the separation of rotations and internal motions in
  the $n$-body problem.
\newblock {\em Rev. Mod. Phys.}, 69:213--275, 1997.

\bibitem{CiftciWaalkens2012}
{\"U}.~\c{C}ift\c{c}i and H.~Waalkens.
\newblock Phase space structures governing reaction dynamics in rotating
  molecules.
\newblock {\em Nonlinearity}, 25:791--892, 2012.

\bibitem{TKKBR2005}
M.~Toda, T.~Komatsuzaki, T.~Konishi, R.~S. Berry, and S.~A. Rice.
\newblock Geometric structures of phase space in multidimensional chaos:
  Applications to chemical reaction dynamics in complex systems.
\newblock In {\em Adv. Chem. Phys.}, volume 130. Wiley, 2005.

\bibitem{EzraWiggins09}
G.~S. Ezra and S.~Wiggins.
\newblock Phase-space geometry and reaction dynamics near index 2 saddles.
\newblock {\em J. Phys. A}, 42(20):205101, 25, 2009.

\bibitem{Halleretal11}
G.~Haller, T.~Uzer, J.~Palaci{\'a}n, P.~Yanguas, and C.~Jaff{\'e}.
\newblock Transition state geometry near higher-rank saddles in phase space.
\newblock {\em Nonlinearity}, 24:527--561, 2011.

\end{thebibliography}

\def\cprime{$'$}

\end{document}